# Tunnel current in self assembled monolayers of 3-mercaptopropyltrimethoxysilane**

Dinesh K. Aswal[1,2,*], Stephane Lenfant[1], David Guerin[1], Jatinder V. Yakhmi[2] and Dominique Vuillaume[1]

[1] *Institut d'Electronique, Microelectronique et Nanotechnologie – CNRS*
*"Molecular Nanostructures & Devices" group*
*BP60069, avenue Poincare, F-59652 cedex, Villeneuve d'Ascq, France*

[2] *Technical Physics and Prototype Engineering Division*
*Bhabha Atomic Research Center, Trombay, Mumbai 400 085, India*

Note and Figure for Table of Content

The current density–voltage (*J-V*) characteristics of self assembled monolayers of 3-mercaptopropyltrimethoxysilane (MPTMS) chemisorbed on the native oxide surface of p[+]-doped Si demonstrate the excellent tunnel dielectric behavior of organic monolayers down to 3 carbon atoms.

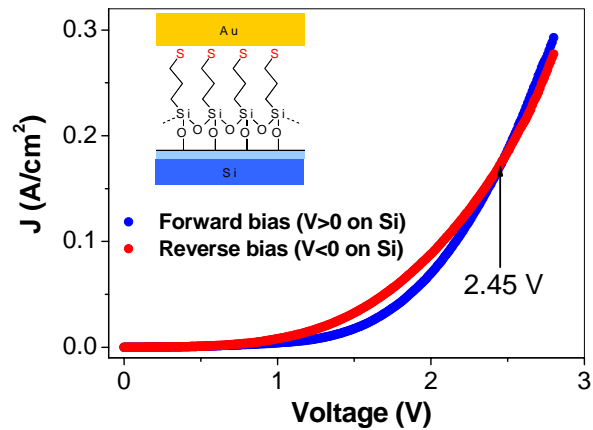

\*       To whom all correspondence should be made; Email: dkaswal@yahoo.com

\*\*      This work is supported by Indo-French Centre for the Promotion of Advanced Research (IFCPAR) through Project 3000-IT-1, by the "Institut de Recherche sur les Composants logiciels et matériels pour l'Informatique et la Communication Avancée" (IRCICA) and by the "ministère de la recherche" through the "AC nanosciences" program.

In future, organic molecules owing to their size, mechanical flexibility and chemical tunability are expected to play a major role in molecular electronic devices - a field termed as "molecular electronics".[1] To realize this expectation, various groups around the world are engaged in finding new methods to measure the electrical properties of the single molecule or a group of molecules. These methods include (i) by stringing molecules in a metal break-junction,[2,3] (ii) by contacting scanning tunneling microscope (STM) or conducting atomic force microscope (CP-AFM) tip to molecular films prepared on a conducting substrate,[4] (iii) by sandwiching molecules between two metal layers through a small orifice or nano-pore[5] and (iv) by adopting "planar sandwich geometry" in which the metal counter-electrodes are directly deposited on self assembled monolayers (SAMs)[6] (see also a review in reference [7]). Working with "planar sandwich geometry" is considered advantageous for molecular electronic applications as one can have ~$10^{15}$ molecules/cm$^2$, and conveniently fabricate 2- and/or 3-terminal molecular devices by choosing appropriate electrodes. At present, two types of SAMs are studied: (i) alkanethiol monolayers on gold substrates and (ii) attachment of organic molecules to SiO$_2$/Si or Si substrates by silanization. Our group has been working for the last several years on the latter approach [8-10] as it has distinct advantages, viz. the possibility of designing new class of resonant tunneling devices[11] owing to an energy gap offered by Si, and the availability of an already existing powerful silicon-based IC industry that can be used effectively for the development of integrated molecular devices.

Our earlier studies on SAMs of n-alkyletrichlorosilanes (n>8 carbon atoms) indicated that the leakage current through them is remarkably low, and thus could be used as dielectric in organic thin film transistors (OTFT).[8-10] Later on, SAM dielectric films were indeed successfully employed for the fabrication of OTFT's by us[12] as well as by other groups.[13] In the present work, we demonstrate that these excellent tunnel dielectric behaviors are still valid down to 3 carbon atoms. Recently, we observed a rectifying behavior when a conjugated (π) group was attached to the alkyl-chain (σ).[14] The rectification ratio of these σ−π SAMs at 1 Volt was as high as 37. However, the maximum attainable current was ~3 x $10^{-5}$ A/cm$^2$, which implies that if we wish to design a σ−π−σ SAM for a resonant tunnel diode the current would be too low to be measured owing to the doubling of the σ chain. In view of this, we have undertaken the synthesis of small alkyl-chain based SAMs.

To begin with, we have chosen 3-mercaptopropyltrimethoxysilane (MPTMS), SH-(CH$_2$)$_3$-Si-(OCH$_3$)$_3$, considering its several advantages (i) it contains the aliphatic chain of only three carbons and has a very short length (0.8 nm), (ii) the end of the molecule is terminated with thiol (-SH) group, which not only would prevent diffusion of counter electrode Au owing to strong chemical bonding between S and Au, but would also enable us to further attach π and/or π−σ moieties for suitable molecular devices, and (iii) the deposition process of SAMs from this molecule is well established, as it is widely used in the controlled covalent immobilization of biomolecules such as DNA and proteins.[15,16]

Here we present the first ever electrical transport measurements made on MPTMS SAMs on Si, which were deposited by a simple vapor phase method. The analyses of current density–voltage (*J–V*) characteristics show that the mechanism of charge transport in this molecule is through tunneling.

The SAMs of MPTMS molecules were synthesized by a simple vapor-phase deposition technique. For this, the source (100 µl of MPTMS kept in a petri-dish) was kept in a vacuum-desiccator and the substrate was mounted upside-down at a distance of 10 cm. The desiccator was evacuated to a vacuum of 0.2 Torr using a dry mechanical pump, and then sealed at this vacuum for ~10 h to facilitate the deposition of SAM on the substrate. The substrates employed were degenerated (100) p-silicon (resisitivity of ~$10^{-3}$ Ωcm), purposely selected to avoid any voltage drop in the substrate during the electrical measurements. The Si substrates had a coverage of native oxide of 1.2 – 1.4 nm (measured by ellipsometry). Prior to the SAM deposition, the Si substrates were cleaned using Piranha solution ($H_2SO_4/H_2O_2$ (V/V) 2:1), and rinsed by de-ionized water.

The thickness of the MPTMS SAMs determined using ellipsometry was 0.8±0.1 nm, which is close to the theoretical value of the length of MPTMS molecule (0.77 nm). The water contact angle measured on the SAMs was 65±3º, which is close to that reported for highly organized MPTMS SAMs[15] and is indicative of –SH group at the outer surface. The atomic force microscopic studies revealed an average surface roughness of 0.14 nm, a value nearly identical to that measured for the native oxide surface of Si, indicating a comprehensive coverage of the monolayer film. The X-ray photoelectron spectroscopic data measured at different take-off angles (i.e. angle between the analyzer acceptance plane and the sample surface) indicated an increase in S:C ratio from 0.19 to 0.23 as the angle is decreased from 70 to 25º. An increase in sulfur content with decreasing take-off angle confirms that the –SH group is at the outer surface and the silane group is attached to the native silicon oxide.

For electrical measurements, 20-80 nm thick gold counter-electrodes of different areas (4 x $10^{-2}$ $cm^2$, $10^{-2}$ $cm^2$, 3.6 x $10^{-3}$ $cm^2$ and $10^{-4}$ $cm^2$) were directly deposited onto the monolayer by thermal evaporation (base vacuum ~$10^{-7}$ Torr and deposition rate = 1Å/s). The measurements of *J–V* characteristics were made with a micromanipulator probe station and HP4140B analyzer (sensitivity 0.1 pA). The *J–V* curves are recorded by applying the bias on Si while Au electrode was grounded. In all, 7 samples were studied, and on each sample 5-10 curves were recorded by using Au counter-electrodes of different areas and thickness. All *J-V* data were reproducible within a range of ± 15 %.

Before presenting our results, we would like to make two points. First, to describe electrical transport in SAMs or in single molecules, there are four possible conduction mechanisms known till date,[17] *viz.* non-resonant tunneling (i.e. direct and Fowler-Nordhiem tunneling), resonant tunneling (through the molecular orbitals in case of molecules with relatively small HOMO-LUMO gap such as π conjugated ones), thermionic conduction and hopping conduction. The tunneling mechanisms are temperature independent but depend on the bias range i.e. direct tunneling takes place at lower bias and a crossover to Fowler-Nordhiem tunneling takes place at higher bias. On the other hand, both thermionic and hopping conduction mechanisms depend on temperature. Thus, in order to compare the *J–V* data with theory, it is essential that the data be taken over a large voltage range as well as their temperature dependence measured. In the present study we have measured *J–V* data in the voltage range -2.8 V to +2.8 V at different temperatures. Second, in our case the thickness of SAM is 0.8 nm, which is much shorter than that of the native oxide of silicon (1.2-1.4 nm). Therefore, it is essential to find the contribution of the native oxide layer on the measured data. For this, we have made 20-80 nm thick Au counter-electrodes directly on the native oxide and measured the *J–V* data. It was observed that compared to MPTMS monolayers (see below) the measured *J* was higher by more than 3 orders of magnitude and, for bias greater than 0.8±0.2 V, the *J–V* curves became linear i.e. ohmic. For instance, at +1.5 V, the *J* values for bare native oxide (thickness 1.2 – 1.4 nm) and with MPTMS SAM, measured from more than 20 junctions each, are respectively, 100±30 and 0.025±0.004 A/cm$^2$. The measured very high *J* value for native oxide is nearly same as that reported in literature, which is attributed to the electrical breakdown owing to the large defects; and in fact is a matter of worry for using it as a dielectric for nanoscale electronics.[18] Thus, the *J–V* data measured on MPTMS monolayer can be treated as its genuine characteristics. Nevertheless the contribution from the native oxide layer particularly at low voltages cannot be treated negligible and thus we have taken into account the thickness of native oxide layer in our calculations.

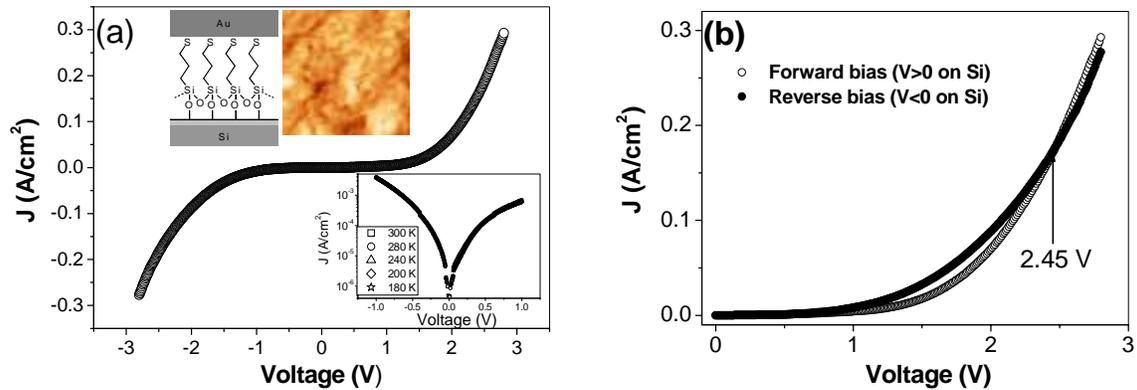

Fig. 1 (a) Typical current density – voltage curve recorded for the MPTMS SAM at room temperature. Upper-left inset: schematic of the Si/SiO$_2$- MPTMS - Au structure and, 200 X 200 nm AFM image of top gold electrode. Bottom-right inset: *J–V* curves in the semi-log scale in ± 1 V range at different temperatures. (Note that all the curves are identical at different temperatures, and were also independent of heating or cooling schedules.). (b) The plots of absolute values of *J* and *V* in reverse as well as forward bias. Note a crossover at 2.45 eV.

A typical *J–V* curve for the MPTMS monolayer measured at 300 K is given in Fig. 1 (a). Two major inferences can be drawn from this figure. (i) The *J–V* curves are asymmetric, which becomes clear when we plot both forward (V>0 on Si) as well as reverse bias (V<0 on Si) data in the same quadrant (Fig. 1(b)) and (ii) The *J–V* data are independent of temperature, which indicate as discussed above, that tunneling is the mechanism of the conduction in this SAM.

From Fig. 1(b) and from *J–V* plots as a function of temperature, it is evident that at low bias the reverse current is higher than the forward one. The rectification ratio (defined as RR = $J_{-1.2 V}/J_{+1.2 V}$), measured from 50 junctions of 7 samples, was found to vary between 2.2 and 5.7. However, at bias greater then 2.45 eV, the direction of rectification is reversed i.e. the forward current is higher than the reverse current. This result is not unexpected, and such an effect has been predicted theoretically by Simmons, way back in 1963,[19] for a thin insulating film sandwiched between two metal electrodes of different work functions i.e. metal(1)-insulator-metal(2) ($M_1$-I-$M_2$) junction. In the present case, the work functions of two electrodes (i.e. p-Si and Au) are, respectively, $\Psi_1$= 4.9 and $\Psi_2$= 5.3 eV.[20] Thus, a qualitative comparison with Simmons theory indicates that our MPTMS monolayer has a very good insulating behavior, and tunneling is the conduction mechanism. In the following, we analyze the *J–V* data of figure 1 using Simmons's equations.

Simmons has shown that in a $M_1$-I-$M_2$ junction, the difference in work functions between two electrodes i.e $\Delta\Psi = \Psi_2 - \Psi_1$ produces an asymmetric potential barrier of a trapezoidal shape. If $\phi_1$ and $\phi_2$ are the barrier heights at the two interfaces of the insulator, then $\Delta\phi$ (i.e. $\phi_2 - \phi_1$) is equal to $\Delta\Psi$ and leads to an asymmetric *J-V* curves with respect to the applied bias. In order to determine the values of $\phi_1$ and $\phi_2$, we have considered the theoretical relationship between *J* and *V* in a tunnel junction having similar electrodes, which is given by

$$J = \frac{\alpha}{d^2}\left[\phi e^{-Ad\sqrt{\phi}} - (\phi + eV)e^{-Ad\sqrt{\phi + eV}}\right] \quad (1)$$

where $\alpha = \frac{e}{4\pi^2\beta'^2\hbar}$ and $A = 2\beta'\sqrt{\frac{2m^*}{\hbar^2}}$ (*e* is the electron charge, $m^*$ is the effective mass of electron in the insulator, $\beta'$ is a constant and has value ~1, $\hbar$ is the reduced Planck's constant), $\phi$ is average barrier height, *d* is the barrier width and *V* voltage between the electrodes. The fit of experimental *J-V* data in reverse and forward bias using equation 1 would thus yield the values of $\phi_1$ and $\phi_2$, respectively.

It has been shown that at *very small voltages* i.e. for eV<<$\phi$, or in the vicinity of V ~ 0, the equation 1, reduces to[19]

$$J = \frac{\gamma\sqrt{\phi}}{d}e^{-Ad\sqrt{\phi}}V \quad (2)$$

where $\gamma = \dfrac{e\sqrt{2m^*}}{4\beta'^2 \pi^2 \hbar^2}$.

Since eV<<$\phi$, it can be considered that $\phi$ does not depend on *V*. Therefore, in this case the *J* is proportional to *V*. As shown in Fig. 2 (a), the data in the bias range ±0.15 V fits well to a linear equation. The slope of linear fit yields a resistance of 4850 ± 30 $\Omega cm^2$ for the MPTMS monolayer, as the bare native oxide layer has a resistance lower by three orders of magnitude.

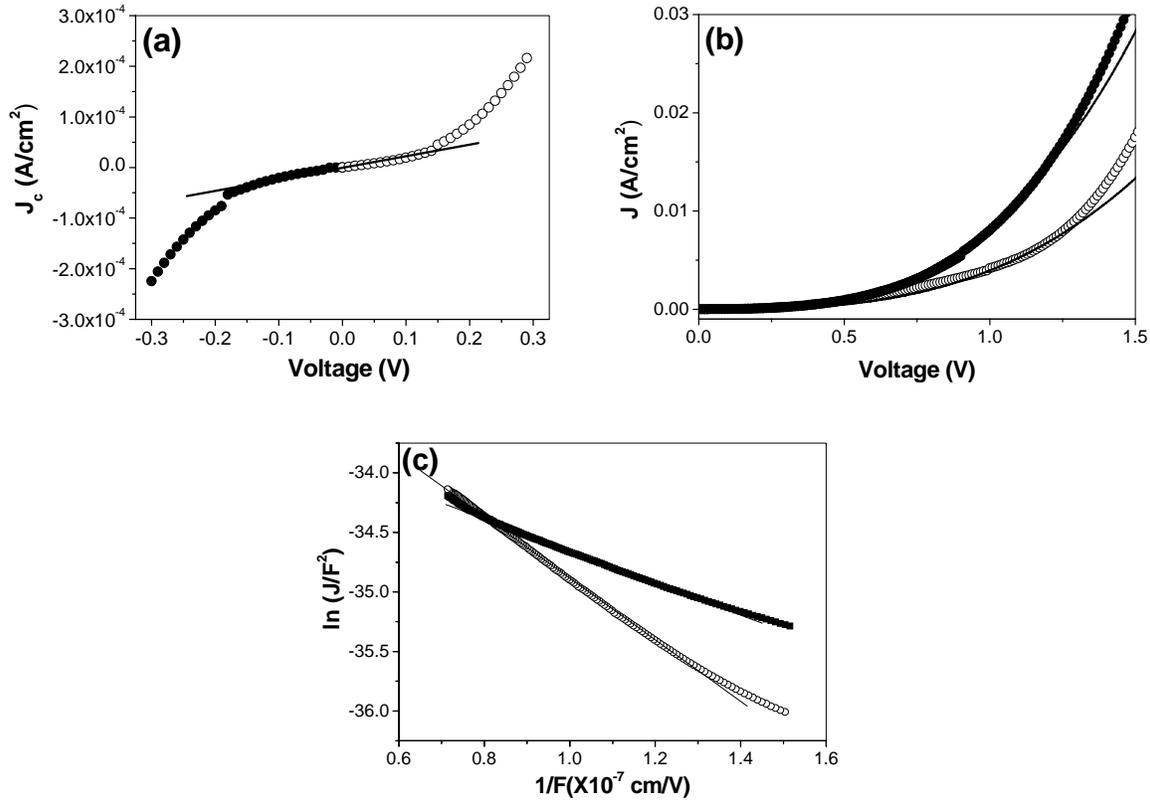

Fig. 2  Data of Fig. 1(a) replotted in three voltage regions (filled circles represent to reverse bias data and open circles to forward bias). (a) Low voltage range i.e. ±0.3 V. The linear fit of data (shown by full line) in ±0.15 V bias range is in accordance with equation 2. (b) Intermediate voltage range. The full curves are fit of data using equation 3 with parameters $m^* = 0.16\ m_e$ ($m_e$ = mass of the electron), $d$ = 2 nm (length of MPTMS molecule + thickness of native $SiO_2$) but with different $\phi$ values in the reverse ($\phi_1$ = 2.14±0.01 eV) and forward ($\phi_2$=2.56±0.01 eV) bias. (c) High voltage range data (i.e. >1.3 V) is plotted as $ln(J/F^2)$ as a function of $1/F$. A linear fit of the data indicate that current transport is dominated by Fowler-Nordhiem tunneling.

For *intermediate voltages* i.e. $eV < \phi/2$, the equation 1 is expressed as [19]

$$J = \frac{\gamma\sqrt{\phi}}{d} e^{-Ad\sqrt{\phi}}(V + \sigma V^3) \tag{3}$$

where $\sigma = \frac{(Ae)^2}{96\phi d^2} - \frac{Ae^2}{32d\phi^{3/2}}$.

The data in the voltage range 0-1.3 V, as shown in Fig. 2(b), fit well to equation 3 using parameters $m^* = 0.16 m_e$ ($m_e$ = mass of the electron), $d = 2$ nm (length of MPTMS molecule + thickness of native SiO$_2$) but with different $\phi$ values in the reverse and forward bias. The fitting of reverse bias data yielded $\phi_1 = 2.14 \pm 0.01$ eV, which corresponds to an electron energy barrier height at MPTMS/Si interface; while an electron energy barrier of $\phi_2 = 2.56 \pm 0.01$ eV is obtained at MPTMS/Au interface by fitting the forward bias data. The difference in barrier heights (i.e. $\Delta\phi = 0.42$ eV) is nearly equal to the difference in the work functions of the two electrodes, which is in accordance with Simmons theory. The value of $m^* = 0.16 m_e$ obtained from fitting is same as experimentally reported by others [21] and close to $m^* = 0.28\ m_e$ theoretically calculated for n-alkanes[22]. With these values of $m^*$ and $\phi$, we deduce a tunneling decay factor ($\beta = A\phi^{0.5}$ in eq. 2) of ~0.65 Å$^{-1}$, which is within the range of 0.5-0.75 Å$^{-1}$ determined experimentally by several groups (see ref. [7]).

Now we move on to *high voltage range* i.e. $eV > \phi/2$. In this case, the potential barrier takes a triangular shape and $\Delta\phi$ causes an asymmetry in the effective barrier width ($d_{eff}$).[19] That is $d_{eff}$ is higher for reverse bias ($d_{eff} = d\phi_1/(eV - \Delta\phi)$) as compared to the forward bias ($d_{eff} = d\phi_2/(eV + \Delta\phi)$); of course $d_{eff} < d$ in both the cases. This leads to a sharper increase of $J$ with $V$ in the forward bias in comparison to the reverse bias and, thus explains a crossover at 2.45 V. In other words, it is well known that a higher Fowler-Nordhiem (FN) current is obtained for a positive bias at the electrode with the smaller tunnel barrier than for the opposite situation, i.e. a positive bias at the electrode with the higher tunnel barrier. In this voltage range the equation 1 reduces to the usual Fowler-Nordhiem form [19] i.e.

$$J = BF^2 \exp(-\frac{C}{F}) \tag{4}$$

where $F = \frac{V}{d}$ is the field across the monolayer, $B = \frac{e^3}{16\pi^2 \hbar m^* \phi}$ and $C = \frac{4\sqrt{2m^*}}{3e\hbar}\phi^{3/2}$.

To test the experimental data in this region, in Fig. 2(c) we have plotted $ln(J/F^2)$ as a function of $1/F$ using $d = 2$ nm. A good linear fit of the data (for $|V| > 1.3$ V)) confirms that at higher voltages the mechanism of conduction in Si/SiO$_2$-MPTMS-Au structure is due to the Fowler-Nordhiem tunneling. From the slopes (C) of the reverse and forward bias data and using $m^* = 0.16 m_e$, the calculated values of $\phi_1$ and $\phi_2$ are, respectively, 0.62 ±0.02 eV and 0.98 ±0.02 eV. These values are much lower than those obtained from the analyses of the intermediate bias range data. However, a better agreement ($\phi_1$ = 1.30 ±0.02 eV and $\phi_2$ = 1.96 ±0.02 eV) is observed when $d$ is taken as 0.8 nm (the length of MPTMS molecule), which indicates that at high voltages the contribution from native silicon oxide layer is insignificant. The remaining discrepancy can be eliminated by taking the image force potential into account, which rounds off the corner and reduces the effective potential barrier height. This effect is expected to be more pronounced in the present case as the thickness of the monolayer is very small and, thus needs further investigations.

The tunneling energy barriers found here, $\phi$ ~2-2.5 eV, are lower than the values, ~4-4.5 eV, that we had previously measured and theoretically calculated on Si/SiO$_x$/alkylsilane/Al junctions with longer chains (≥ 12 carbon atoms).[9] As discussed by Salomon et al[7], the nature of electrode contacts with monolayer does affect the barrier height for tunneling e.g. the chemical bonds with electrodes would decrease the HOMO-LUMO gap of the alkyl chains (~9 eV for long alkyl chains, and calculated even larger for small chains [9]) and in turn lower the barrier height. Thus, our earlier high $\phi$ values could be attributed to the poor contacts between the terminating methyl group of the monolayers and Al; while lower $\phi$ values obtained in the present study is due to the fact that the thiol ends chemically react with the first evaporated gold atoms forming S-Au bonds at the upper interface (a comprehensive coverage of Au on the monolayer is evident from the AFM image shown in the Fig. 1 (a)). We can also note that a 2-2.5 eV value for Si/SiO$_x$/alkyl SAM/metal when the alkyl molecule form a chemical bond at the metal side is also consistent with the value found for metal/alkylthiol(or Dithiol)/metal where the molecule is similarly chemisorbed at the metal surface.[7,21,23-25] The present study also yields an effective mass of electrons, $m^* = 0.16\ m_e$, which is somewhat closer to $m^* = 0.28\ m_e$ theoretically calculated for n-alkanes[22].

In conclusion, the *J-V* characteristics of MPTMS SAMs on Si are found to be asymmetric, and the direction of rectification has been found to depend upon the applied voltage range. At voltages < 2.45V, the reverse bias current was found to be higher than forward bias current; while at higher voltages this trend was reversed. This result is in agreement with Simmons theory. The tunnel barrier heights for this short chain (2.56 and 2.14 eV respectively at Au and Si interfaces) are in good agreement with the ones for longer chains (>10 carbon atoms) if the chain is chemisorbed at the electrodes. These results extend all previous experiments on such molecular tunnel dielectrics down to 3 carbon atoms. This suggests that these molecular monolayers, having good tunnel behavior (up to 2.5 eV) over a large bias range, can be used as gate dielectric well below the limits of Si-based dielectrics. The present study also suggests that interpreting asymmetric *J-V* curves could be deceptive if one does not analyze the data in a wide bias range as well as measure their temperature dependence.